\documentclass[12pt]{article}
\usepackage{epsfig}

\begin{document}


\def\a{\alpha}
\def\b{\beta}
\def\g{\gamma}
\def\G{\Gamma}
\def\d{\delta}
\def\D{\Delta}
\def\e{\epsilon}
\def\h{\hbar}
\def\ve{\varepsilon}
\def\z{\zeta}
\def\t{\theta}
\def\vt{\vartheta}
\def\r{\rho}
\def\vr{\varrho}
\def\k{\kappa}
\def\l{\lambda}
\def\L{\Lambda}
\def\m{\mu}
\def\n{\nu}
\def\o{\omega}
\def\O{\Omega}
\def\s{\sigma}
\def\vs{\varsigma}
\def\S{\Sigma}
\def\vphi{\varphi}
\def\av#1{\langle#1\rangle}
\def\pa{\partial}
\def\na{\nabla}
\def\hg{\hat g}
\def\un{\underline}
\def\ov{\overline}
\def\cF{{\cal F}}
\def\cG{{\cal G}}
\def\cN{{\cal N}}
\def\Hsl{H \hskip-8pt /}
\def\Fsl{F \hskip-6pt /}
\def\cFsl{\cF \hskip-5pt /}
\def\ksl{k \hskip-6pt /}
\def\pasl{\pa \hskip-6pt /}
\def\tr{{\rm tr}}
\def\tcF{{\tilde{{\cal F}_2}}}
\def\tg{{\tilde g}}
\def\shalf{\frac{1}{2}}
\def\nn{\nonumber \\}
\def\w{\wedge}
\def\ra{\rightarrow}
\def\la{\leftarrow}
\def\be{\begin{equation}}
\def\ee{\end{equation}}
\newcommand{\brr}{\begin{eqnarray}}
\newcommand{\err}{\end{eqnarray}}

\begin{titlepage}
\begin{flushright}    
hep-ph/9911439
\end{flushright}
\vskip 1cm
\centerline{\LARGE{\bf Gauge mediated supersymmetry breaking}}
\centerline{\LARGE{\bf and supergravity}}
\vskip 1.5cm
\centerline{\bf 
S. P. de Alwis \footnote{e-mail address: dealwis@pizero.colorado.edu} and 
Nikolaos Irges \footnote{e-mail address: irges@pizero.colorado.edu}}
\vskip .7cm
\centerline{\em Department of Physics, Box 390}
\centerline{\em University of Colorado}
\centerline{\em Boulder, CO 80309, USA}
\vskip .2cm
\vskip 1.5cm
\centerline{\bf {Abstract}}

We analyze simple models of gauge mediated supersymmetry breaking in the
context of supergravity. We distinguish two cases. One is when 
the messenger of the supersymmetry breaking 
is a non Abelian gauge force and the other is when the messenger
is a pseudoanomalous $U(1)$. 
We assume that these models originate from string theory and
we impose the constraint of the vanishing of the cosmological 
constant requiring also the stabilization of the dilaton.
In the first case, we do not find vacua that are consistent 
with the constraints of gauge mediation and have a zero 
tree level cosmological constant.  
In the second case, no such conflict arises.
In addition, by looking at the one
loop cosmological constant, we show that the
dilaton $F$-term can not be neglected in either case.
For the gauge mediated case our considerations suggest 
that the dilaton must be frozen
out of the low energy field theory by non-perturbative string dynamics.
\vskip .5cm       
\vfill
\end{titlepage}

\section{Introduction}
Spontaneously broken global supersymmetric theories, 
generate a positive vacuum energy making a flat space 
solution inconsistent. As is  well known the proper 
framework for discussing spontaneous breaking of 
supersymmetry is supergravity. The natural scale for such
a theory is however the Planck scale $M_{Pl}$, which is  many 
orders of magnitude (at least 15) larger than the 
expected SUSY breaking splitting between standard 
model particles and their superpartners.
The question then arises as to how this wide 
separation of scales can be generated. In gravity 
mediated supersymmetry breaking schemes an
intermediate scale is typically generated 
by the condensation of 
gauginos in some hidden sector gauge group\cite{Nilles}. 
If $S$ is the dilaton, a
possible superpotential for the hidden sector could thus take the 
form
\be W=M_{Pl}^3e^{-{S\over b}}+kM_{Pl}^3\ee
with a constant $k$ in the superpotential whose 
origin is not understood at this point and $b$ a model dependent
parameter.
\footnote{Another popular model, the so-called 
race track one where two or more gauge groups 
condense does not need a constant but appears 
to be ruled out on other grounds \cite{BA}.}
Such a super potential can be used to stabilize the dilaton and
give SUSY breaking at an intermediate scale 
$\L  =\sqrt{m_{3/2}M_{Pl}}$ with zero cosmological 
constant. The latter in fact involves cancellation 
between two terms both of which are of order $\L^4$ and 
thus can be accomplished by fine tuning a number 
of order one. Generically however such gravity 
mediated SUSY breaking models will
lead to large flavor changing neutral currents, unless the $F$ term of
the dilaton dominates the SUSY breaking
\footnote{If the SUSY breaking is dilaton 
dominated this could be avoided but it is fair 
to say that there is no natural framework 
coming from string theory in which this can be realized.}. 
 
There are two other scenarios that have been 
discussed in the literature that 
lead to SUSY breaking, such that no 
large flavor changing neutral currents are 
generated, both of which communicate the 
SUSY breaking from a hidden sector to the 
visible by means of gauge interactions rather than by gravity.
One, is when the information of supersymmetry breaking 
is communicated to the visible sector by the
usual gauge (non-Abelian) interactions \cite{GuidRat}. 
This mechanism is called Gauge
Mediated Supersymmetry Breaking (GMSB), an old idea 
recently revived in \cite{DN}.
The quantity that sets the scale for the soft masses 
is ${\alpha \over {4\pi}}{F_X\over <X>}$, where
$F_X$ and $<X>$ are the highest and lowest components of a 
chiral superfield $X$ in
the hidden sector. Phenomenological reasons 
\footnote{One of the constraints on $<X>$ comes from standard Big Bang
nucleosynthesis, giving an upper bound of $<X>\le 10^{12}$ $GeV$ 
\cite{Dimopoulos}.}
require $<X>$ to be rather
low compared to the cut-off scale $M\sim M_{Pl}$.

In contrast to the gravity mediated case though, in these models the 
cancellation of the cosmological constant 
is rather artificial. In fact, at the minimum the scalar potential we need,
\be V_0 \simeq |F|^2-3{|W|^2\over M_{Pl}^2}=0.\ee
If the supersymmetry breaking scale is 
$\sqrt{F}\sim \L_{SUSY}$ then to satisfy this we 
need a constant in the superpotential 
$W\sim M_{Pl}\L_{SUSY}^2$ which  is much 
larger than the SUSY breaking scale. Thus  
for instance for $\sqrt{F}\sim 10^5 GeV$ as is typically the case in this 
scenario, then a constant at a much larger 
scale $\sim 10^{9}GeV$ needs to be added. 

One may generate such a constant by adding yet 
another gauge sector to the model \cite{Yanagida}. 
Thus one possiblity is to add a SUSY
preserving SQCD sector, which will generate a 
non-perturbative contribution to the superpotential 
giving for instance an effective additional piece,
\be W_{eff}=\L^{3N_c-1\over N_c-1}
(Q\overline{Q})^{-{1\over N_c-1}}+\l Q\overline{Q}
\tilde{S}-{g\over 3}\tilde{S}^3\ee 
where the gauge theory is $SU(N_c)$ and $Q,\overline{Q}$ are quark, anti quark superfields.
The total superpotential is then
$W_{tot}=W(X,...)+W_{eff}$. The SUSY preserving 
conditions $F_{\tilde S}=F_Q=F_{\overline Q}=0$ then give 
${\tilde S}\sim \sqrt {(Q\overline{Q})}\sim\L$ giving 
effectively a constant of the order of
$\L^3$ in the superpotential. Now the question arises as to how this
scale is related to the fundamental scale of supergravity, namely
$M_{Pl}$. In fact one should expect that $\L =e^{-S\over 3}M_{Pl}$
where $S$ is the four dimensional dilaton as 
in the above discussion of gravity mediated 
SUSY breaking. But now it is clear that without
adding a constant to the super potential, $S$ 
will have runaway behaviour and we 
will not be able to generate an intermediate scale $\L$.

In this paper we will not worry about the origin of such a constant
and will leave that to some unknown (stringy) 
mechanism as in the case of the gravity 
mediated model. We will simply add such a 
constant but ask whether whithin the assumptions 
of the gauge mediated suspersymmetry breaking 
model, one could stabilize the dilaton in the same 
way that one is able to do in the supergravity 
mediated models. The answer will turn out to be negative. 
This does not imply that this mechanism is not viable. But it suggests
 (given the remarks in the previuos paragraph) 
that the dilaton 
(and presumbaly all moduli) of string theory are 
stabilized by some nonperturbative stringy mechanism, 
and that it is frozen out of the low energy field theory 
dynamics. Indeed given the difficulties of finding 
viable mechanisms for the gravity mediated case 
this may be required in that case also 
(see for example the third paper of \cite{BA}).

The other scenario, is when supersymmetry breaking is communicated to the
visible sector by a pseudoanomalous $U(1)$ gauge interaction (U1MSB)
\cite{BD}. 
The scale of the soft masses in this case is set either 
by the $D$-term corresponding to the $U(1)$ or, in the absence of
$D$-term contributions, it is set by an $F$-term.
Since the virtue of the pseudoanomalous $U(1)$ is mainly the generation
of fermion masses, it is assumed to be flavor non universal over the
visible sector. Then, however, in order to avoid conflict with data on
flavor changing neutral currents (fcnc), one has to require that the
$D$-terms essentially do not contribute to supersymmetry breaking.
\footnote{This can be achieved in some models with anomalous $U(1)$,
under additional assumptions.}
Furthermore, to have universal masses at the scale $M$, (again for fcnc
reasons) one would like, in addition, supersymmetry breaking 
to be dominated by the dilaton $F$-term, so that the superpartners  
would get universal masses $\sim F_S/M$.  
We will analyze prototype models for
each case and we will try to answer the question under what
circumstances supersymmetry is unbroken or broken 
with vanishing cosmological constant and what are the main
features of the corresponding unbroken or broken vacua. 
 \par 
For GMSB, we will take the tree level superpotential to be simply 
$w_0=\lambda X{\Phi}{\bar \Phi}$, with $\lambda $ a Yukawa coupling.
$\Phi$ and $\bar \Phi$ are the messenger fields,
vector-like with respect to the gauge group that contains the standard
model gauge group. $X$ is a field, singlet of the visible and hidden
sector gauge groups. For concreteness, we will take a specific,  
$SU(N_c)$ hidden sector that contains 
one family ($N_f=1$) of fields $Z$ and ${\bar Z}$ 
transforming respectively as ${\bf N}$ and ${\bf \bar N}$ of $SU(N_c)$.
If the hidden sector is
asymptotically free, then at the confining scale the formation of
gaugino condensates becomes possible.
Below this scale, the relevant physical hidden sector field becomes the
hidden sector singlet ``meson'' $X\equiv \sqrt{Z{\bar Z}}$,
and the corresponding contribution to the superpotential,
for $N_c>N_f+1$, will be
\cite{ADS} $w(S,X)=c\cdot X^{-p}e^{-rS}$.
Here $c$, $p$ and $r$ are model dependent 
parameters and $S$ is the dilaton field. Given that 
we assumed one hidden sector flavor, 
$p=2/(N_c-N_f)$ and $r=8\pi^2/(N_c-N_f)$ are both
strictly positive quantities. 
This specific choice of the hidden sector, will not spoil the generality
of our final conclusions concerning supersymmetry breaking and the vanishing
of the cosmological constant. 
The full superpotential is then
${\cal W}=w_0(X,{\Phi},{\bar \Phi})+w(S,X)+k$. 
In ${\cal W}$ we allow for a constant term $k$ which can be either the
vacuum value of a combination of fields that have been set to constants
upon minimization of the scalar potential with respect to the
corresponding fields, or a new, non perturbative contribution to
${\cal W}$. In either case what is important for us is that $k$ is
independent of $S$, $X$, ${\Phi}$ and ${\bar \Phi}$.
\par For U1MSB, we assume a hidden sector which has exactly the same
structure as the hidden sector of the GMSB model and we 
couple it to a visible sector
singlet field ${\Theta}$. Below the condensation scale, the
physical degree of freedom is $\sqrt{{Z}{\bar Z}}\equiv X$.
Both $\Theta$ and $X$ are assumed to be charged under the $U(1)$. 
The full superpotential in this case is taken to be
${\cal W}=w_0+w+k=\lambda \Theta X^2+c\cdot X^{-p}e^{-rS}+k$, where
$k$ is independent of $\Theta$ and $X$
\par Due to the dilaton dependence of $w$, 
it is naturally implied that these
superpotentials arise from some string compactification which requires
supergravity as the correct low energy effective theory. 
In supergravity, the scalar potential 
(ignoring $D$-terms), is given by \cite{sugra}:
\be V=e^{K/M^2} \Bigl[\sum _{\phi _i}{|F_{\phi _i}|^2}
-3{|{\cal W}|^2\over M^2}\Bigr] \;\;\;\;
{\rm with} \;\;\;\; F_{\phi_i}\equiv
{\cal W}^{(\phi_i)}+
{K^{(\phi_i)}\over M^3}{\cal W},\label {eq:V}\ee
where $K$ is the K$\ddot{\rm a}$hler potential,
and $\phi_i$ denotes any of the physical fields.
Superscripts in parentheses denote differentiation with respect to
the corresponding fields.
We will assume for simplicity that all the other moduli besides the dilaton
(such as $T$ or $U$) can be neglected.
The K$\ddot{\rm a}$hler potential $K$ then will simply be:
\be K=-{M^2}\log{({S}+{\bar S})}-
\sum _{\phi _i}\phi_i\phi_i^*\label{eq:Kahler}.\ee  
\par 

 The K$\ddot{\rm a}$hler potential has been assumed to have its
usual tree level value. One could imagine 
adding other, weak coupling or strong
coupling corrections to it, but it seems unlikely that 
such terms would significantly  affect our 
analysis. In fact it should be noted that 
although we are assuming some stringy effect 
which will generate a constant in the 
superpotential this does not mean that the 
four dimensional dilaton $S$ is not in the 
weak coupling region. Indeed, the dilaton being 
related to the gauge coupling, ought to be fixed 
at some weak coupling value. The ten dimensional 
dilaton which governs string perturbation theory 
on the other hand may be at the S-dual point so 
that string nonperturbative effects are important. 
Thus, we believe that we are justified in 
using the weak coupling form of the K${\ddot {\rm a}}$hler potential 
for $S$ and ignoring perturbative and nonperturbative corrections.

\par 
It is also convenient to 
set $M=1$ and to define ${\cal G}\equiv K-\log {|{\cal W}|^2}$.
The task is to minimize the potential ($\ref {eq:V}$) 
for the GMSB and U1MSB models and
see if supersymmetry breaking vacua with desired phenomenological 
properties exist. 
\par In section $2$, we investigate 
supersymmetric vacuum configurations with vanishing 
tree level cosmological
constant. A necessary and sufficient condition for this is:  
\be F_{\phi _i}=0 \; , \;\;\; 
D=0 \;\;\;\;
{\rm and} \;\;\;\; {\cal W}=0.\label{eq:susy}\ee 
The $D=0$ condition is necessary in the $U(1)$ case.
In fact, we will show that there
are no supersymmetric vacua for either case with a nonzero gauge
coupling. In section $3$, we investigate supersymmetry breaking.
To look for supersymmetry breaking vacua, 
we have to minimize the potential.
The scalar potential $V=e^ K [\cdots ]$ upon minimization, gives 
$V^{(\phi _i)}=K^{(\phi _i)}e^K [\cdots]+e^K {[\cdots ]}^{(\phi _i)}=0$, 
which implies
that in a vacuum with vanishing 
tree level cosmological constant it is sufficient
to solve for ${[\cdots ]}^{(\phi _i)}=0$, provided that $K\ne -\infty$ at
minimum. 
In section $4$, we discuss vanishing of the  
cosmological constant at one loop. 
\par
In section $6$, we give our conclusions.

\section{Supersymmetric Vacua}

\centerline{\bf GMSB model}
\par
Let us start by writing out our superpotential again as \cite{GuidRat}: 
\be {\cal W}=
w_0+w+k=\lambda X{\Phi}{\bar \Phi}+cX^{-p}e^{-rS}+k. \label{eq:WGMSB}\ee
Using the  K$\ddot{\rm a}$hler potential ($\ref{eq:Kahler}$),
the $F$-terms are computed to be:
\be F_X={1\over X}(w_0-pw)-X^*{\cal W}\; , \;\;\;\;
F_S=-rw-\sigma {\cal W},\label {eq:FGMSB1}\ee
\be F_{\Phi}={1\over \Phi}w_0-\Phi^*{\cal W}\; , \;\;\;\;
F_{\bar \Phi}={1\over {\bar \Phi}}w_0-
{\bar \Phi}^*{\cal W},\label {eq:FGMSB2}\ee
where we have defined for convenience $\sigma\equiv 1/(S+{\bar S})$.
Since the supergravity framework in which we work is valid only up to
the scale $M$, we will exclude from our analysis cases with the vevs of
$X$, ${\Phi}$, ${\bar \Phi}$ or $Y=\infty$. 
We will denote a field and its vacuum value (vev) by the same symbol,
since there is no possibility of confusion.   
We can distinguish the following possibilities: 
\begin{itemize}
\item ${w\ne 0}$:
The constraints ($\ref{eq:susy}$) allow us to write the dilaton $F$-term
condition as $F_S=w(-r)=0$, which requires $r=0$. But this is not possible,
since we saw that $r$ is a strictly positive quantity. 
\item ${w=0}$. 
\par {\bf 1}. $S\ne \infty$ ($\sigma \ne 0$): 
$w$ can be zero only if $X=0$ provided $p<0$.
However, since $p$ is a
strictly positive quantity, this is not an allowed solution.  
\par {\bf 2}. $S=\infty$ ($\sigma =0$):
For $p>0$, there is a supersymmetric vacuum configuration with
$X\ne 0$ and $\Phi={\bar \Phi}=0$. 
\end{itemize}
We conclude that there is no supersymmetric minimum with 
a vanishing tree level cosmological constant and a finite value
for the dilaton and therefore a nonzero gauge coupling.    

\centerline{\bf U1MSB model}
\par 
The superpotential in this model is:
\be {\cal W}=w_0+w+k=\lambda \Theta X^2+cX^{-p}e^{-rS}+k
\label{eq:WU1MSB}.\ee
With the  K$\ddot{\rm a}$hler potential ($\ref{eq:Kahler}$)
the $F$-terms are:
\be F_{\Theta}={w_0\over \Theta}-\Theta^*{\cal W},\;\;\;
F_S=-rw-\sigma {\cal W}, \;\;\; 
F_{X}={1\over X}(w_0-pw)-X^*{\cal W}.\label {eq:FU1MSB}\ee
Again, we distinguish two possibilities:
\begin{itemize}
\item ${{w\ne 0}}$:
Using ${\cal W}=0$ we can write $F_S=w(-r)=0$ 
and since $r>0$, there is no such supersymmetric vacuum.
\item ${w}$=0:
\par {\bf 1}. $S\ne \infty$ ($\sigma \ne 0$): 
$w$ can be zero only if $\Theta=0$, which requires $p<0$.
Since $p$ is strictly positive, this is not an allowed solution.  
\par {\bf 2}. $S=\infty$ ($\sigma =0$):
The $F_{\Theta}=0$ equation implies $X=0$. There is a supersymmetric vacuum
with $p>0$ and $\Theta$ undetermined.
\end{itemize}
We conclude that in the U1MSB model too,
there is no supersymmetric vacuum except with zero gauge
coupling.      

\section{Supersymmetry Breaking Vacua}

\centerline{\bf GMSB model}
\par
In the GMSB model, there are four types of fields, namely the hidden
sector field $X$, the messengers $\Phi$ and ${\bar \Phi}$ and the
dilaton $S$. The physically relevant case is when the 
standard model nonsinglet messenger fields
do not take vacuum expectation values, i.e. $\Phi={\bar \Phi}=0$, which
implies through equations ($\ref{eq:FGMSB2}$) that
$F_{\Phi}=F_{\bar \Phi}=0$.
The minimization conditions simplify considerably 
if we notice that the values of the fields $X$
and $\sigma$ that minimize the potential with $V=0$, will minimize 
${\tilde V}\equiv {V\over {3|{\cal W}|^2}}$ as well, since
\be {\tilde V}^{(\phi_i)}={{V}^{(\phi_i)}\over {3|{\cal W}|^2}}-
{{V}\over {3|{\cal W}|^4}}(|{\cal W}|^2)^{(\phi_i)},\ee
and
that $w_0$ is essentially absent from the picture due to the vanishing
of ${\Phi}$ and ${\bar \Phi}$. Define then
\be {\tilde F}_a\equiv {F_X\over {\sqrt {3}{\cal W}}}=
{-1\over \sqrt {3}}\Bigl({{p(1-{\tilde k})}\over
X}+X^*\Bigr)={-1\over \sqrt {3}}\Bigl({{p(1-{\tilde k})}
\over a}+a\Bigr)e^{-i\alpha},\label{eq:f1}\ee   
where ${\tilde k}\equiv k/{\cal W}$. 
We derived the second part of the above equation using the definition of
the $F$-term and the expression for ${\cal W}$ and in the third part of
the equation we have defined $X\equiv ae^{i\alpha}$. 
Similarly, we define 
\be {\tilde F}_{S}\equiv {F_{S}\over {\sqrt {3}{\cal W}}}={-1\over \sqrt
{3}}\Bigl({r(1-{\tilde k})}+\sigma \Bigr).\label{eq:f2}\ee 
The phase in the above is zero if $k=0$. These definitions allow us to write
${\tilde V}=|{\tilde F}_a|^2+|{\tilde F}_{S}|^2-1$.   
Now let us look for solutions to the minimization conditions 
with $V=0$ and try to find out
if it is possible to generate the scale hierarchy necessary for
a low energy gauge mediated supersymmetry breaking scenario.  
For simplicity, we will ignore all phases.  
The vacuum conditions become:
\be {\tilde V}^{(a)}: \;\;\; 
{\tilde F}_a\Bigl[1-{p\over a^2}(1-{\tilde k})(1+p{\tilde k})\Bigr]
+{\tilde F}_{S}\Bigl[-{pr\over a}{\tilde k}(1-{\tilde k})\Bigr]=0,
\label{eq:GMVAC1}\ee
\be {\tilde V}^{(S)}: \;\;\; 
{\tilde F}_a\Bigl[-{pr\over a}{\tilde k}(1-{\tilde k})\Bigr]+
{\tilde F}_{S}\Bigl[-r^2{\tilde k}(1-{\tilde k})-\sigma^2\Bigr]=0,
\label{eq:GMVAC2}\ee
\be {\tilde V}=0: \;\;\; 
\Bigl[{p\over a}(1-{\tilde k})+a\Bigr]^2+
\Bigl[r(1-{\tilde k})+\sigma\Bigr]^2=3 \label{eq:VAC}.\ee 
We can rewrite ($\ref{eq:GMVAC1}$) and ($\ref{eq:GMVAC2}$) as
\be {{\tilde F}_a\over {\tilde F}_{S}}=
-{{ar}\over {p}}-{{a\sigma^2}\over {pr{\tilde k}(1-{\tilde k})}}
={{{p\over a}(1-{\tilde k})+a}\over 
{r(1-{\tilde k})+\sigma}}\label{eq:GMVAC3},\ee 
\be \Bigl[1-{p\over a^2}(1-{\tilde k})(1+p{\tilde k})\Bigr]
\Bigl[-r^2{\tilde k}(1-{\tilde k})-\sigma^2\Bigr]+
\Bigl[-{pr\over a}{\tilde k}(1-{\tilde k})\Bigr]
\Bigl[-{{pr}\over a}{\tilde k}(1-{\tilde k})\Bigr]=0\label{eq:GMVAC4}.\ee
The first part of the equation (\ref{eq:GMVAC3}) comes from 
the vacuum condition ${\tilde V}^{(S)}=0$
and the second from the expressions 
($\ref{eq:f1}$) and ($\ref{eq:f2}$) for the $F$-terms.
Having in mind the constraint 
$a\le {\cal O}(\eta)\sim 10^{-6}$ mentioned in the 
introduction and that $r\sim {\cal O}(10)$, $p\sim {\cal O}(1)$,
we first notice that
(\ref{eq:VAC}) can be satisfied only if $(1-{\tilde k})$ is also small, say
$(1-{\tilde k})\sim {\cal O}(\eta \epsilon)$, with $\epsilon$ at most of
order one.
Then, (\ref{eq:GMVAC3}) can be written as
\be {\cal O}\Bigl(\eta r+{\sigma^2\over {r\epsilon}}\Bigr)=
{\cal O}\Bigl(
{{\epsilon +\eta}\over {r\eta \epsilon +\sigma}}\Bigr).\label{eq:R}\ee
We distinguish two possibilities. First, assume that $\sigma \sim {\cal
O}(1)$. Then, (\ref{eq:R}) can be solved if $\epsilon\sim {\cal
O}({1\over \sqrt{r}})$. Using, however, this value for $\epsilon$,
(\ref{eq:GMVAC4}) becomes 
${\cal O}({1\over {\eta \sqrt{r}}})\sim {\cal O}(r)$, which
can not be satisfied since $\eta <<1$.
If  on the other hand, $\sigma <{\cal O}(1)$, 
then from (\ref{eq:VAC}) we see that 
$\epsilon \sim {\cal O}(1)$ and (\ref{eq:GMVAC3}) becomes
\be {\cal O}\Bigl(\eta r+{\sigma^2\over {r}}\Bigr)=
{\cal O}\Bigl({{1}\over {r\eta+\sigma}}\Bigr).\label{eq:R1}\ee
We now turn to (\ref{eq:GMVAC4}), which can be solved only if 
$\sigma ^2\sim {\cal O}(\eta r^2)$. Substituting this back into 
(\ref{eq:R1}), we get the condition 
${\cal O}(\eta r)\sim {\cal O}({1\over {r\sqrt{\eta}}})$, which is not
satisfied for $\eta <<1$. 
\footnote{If we relax the constraint $a\sim \eta$, we can solve the
vacuum equations with $\sigma \sim 1$, and $a$, 
$(1-{\tilde k})\sim 1/r$. This seems to be the only possibility to avoid the
negative conclusions of our analysis.} 
We conclude that we can not satisfy the vacuum equations with 
$a\sim {\cal O}(\eta)$ and a vanishing tree level cosmological constant.       
\par It is natural to ask how general this conclusion is, given that we
have considered only a certain class of superpotentials. More
specifically, we have not considered 
additional fields and couplings possible in the superpotential
involving the field $X$. Such a coupling, can be of the form of
either $XAB$ or $XXA$ or $XXX$, where $A$ and $B$ are chiral superfields
of the hidden sector. 
Clearly, the most important contribution comes from $XAB$,
when $A,B\sim {\cal O}(1)$.
Then, defining $l\equiv <AB>$, 
${\tilde l}\equiv l/{\cal W}$ and ignoring phases, we get the
leading order modification to the $F_a$-term:
\be {\sqrt 3}{\tilde F}_a\rightarrow {\sqrt 3}{\tilde F}_a
+{p\over a}{\tilde l},\ee
which amounts to ${\tilde k}\rightarrow {\tilde k'}$, with 
${\tilde k'}={\tilde k}+{\tilde l}$. By similar arguments as before,
we can see that our previous conclusion remains.

\centerline{\bf U1MSB model}
\par The first question to address is
if the $U(1)$ symmetry is flavor universal in the visible sector or not. 
The main argument for the existence of the pseudoanomalous $U(1)$ is
that, provided that it is flavor dependent, 
it is an excellent candidate to explain fermion mass hierarchies \cite{ILR}. 
But then, after supersymmetry
breaking, its $D$-term will contribute 
to the soft masses which in turn tend to give large flavor changing
neutral current (fcnc) contributions. 
Without any uneven mass splittings between squark generations,
the U1MSB scenario is therefore consistent with data
only if the dilaton $F$-term dominates over the $D$-terms.
Here, we will assume that the $D$-term is negligible 
(under some additional assumptions it can indeed be small \cite{CI})
and argue that
there exists a supersymmetry breaking vacuum with vanishing
tree level cosmological constant.     
\par The most important feature in the U1MSB model is that
we do not have to make any special assumptions about the vevs of any of
the fields. Each minimization condition 
can in principle determine its corresponding field vev. 
Indeed, a solution to the minimization conditions,
as an expansion in the small parameter 
$\epsilon \equiv {\Theta}^2/Y^2$, has been presented in $\cite{BD}$. 
Finally, cancelation of the tree level cosmological constant  
can be achieved by an appropriate choice of $k$.   
Therefore, in this model, the vacuum conditions can
be satisfied and at the same time a supersymmetry breaking scale of
${\cal O}(100)$ $GeV$ can be generated. 
This scenario of supersymmetry breaking, 
in the dilaton dominance limit, qualitatively is very similar 
to gravity mediation. The difference comes from the field $\Theta $
which when taking a vev, provides us with an 
additional supersymmetry breaking parameter.

\section{One loop cosmological constant}
 
To ensure the vanishing of the cosmological constant at one loop, 
one has to look at the one loop effective potential \cite{CW}:
\be V_1=V+{1\over {64\pi^2}}Str {\cal M}^0\cdot M ^4\log
{M^2\over \mu^2}+{1\over {32\pi^2}}Str {\cal M}^2\cdot M ^2+
{1\over {64\pi^2}}Str {\cal M}^4\log {{\cal M}^2\over M^2}, \ee
with 
\be Str {\cal M}^n=\sum_{i}(-1)^{2J_i}(2J_i+1)m_i^n\ee
the well known supertrace formula of supergravity. Clearly, 
to have a zero cosmological constant, it is not
sufficient to set $V=0$ alone.
The second term is zero in theories with equal number of bosons and
fermions and in particular in all supersymmetric models, but
the third and the last term have to be taken into account.
The last term is one that does not destabilize the hierarchy,
but its presence is important for the 
vanishing of the cosmological constant. It can be taken into account
by modifying the constraint on the classical potential $V=0$, to
$V+{1/{(64\pi^2)}}Str {\cal M}^4\log {{\cal M}^2\over M^2}=0$.
The additional term, however, is expected to be negligible compared to
$V$, so we could safely ignore it in the previous section. 
In fact, we used the constant $k$ introduced in the superpotential to carry
out this ``continuous'' ($V=0$) fine tuning.  
The vanishing of the third term on the other hand 
is required in order to have a stable
hierarchy, and it involves a ``discrete'' fine tuning. 
Given that $Str {\cal M}^2=2 Q m_{3/2}^2$,
with $m_{3/2}$ the gravitino mass and 
\be Q=N-1-{\cal G}^iH_{i {\bar j}}{\cal G}^{\bar j},\label{eq:Q}\ee
with
$H_{i {\bar j}}=\partial _{i}\partial _{\bar j}\log \det {\cal
G}_{m\bar n}-\partial _{i}\partial _{\bar j}\log \det {Re [f_{ab}]}$,
a necessary condition for the vanishing of 
the cosmological constant is $Q=0$ \cite{FerKounZwir}.
In the above, ${\cal G}_i={\partial {\cal G}\over {\partial {\phi}^i}}$,
$N$ is the total number of chiral multiplets, 
$f_{ab}=S\delta _{ab}$ is the gauge kinetic function and ${\cal
G}_{m\bar n}$ is the K$\ddot{\rm a}$hler metric of the K$\ddot{\rm
a}$hler manifold for the $N$ chiral superfields.
We can readily calculate $Q$ for both cases. 
${\cal G}_{S}={F_S\over {\cal W}}$, ${\cal G}_{\bar S}={\bar {\cal
G}_{S}}$ and  
the K$\ddot{\rm a}$hler metric is 
${\cal G}_{i \bar j}=diag(1/(S+\bar S)^2,-1)$, where the $-1$ is
to be understood as multiplied by the 
$(N-1) \times (N-1)$ unit matrix,
where $N$ is the total number of chiral superfields.
This is true only if the only modulus in the model is the dilaton
and the K$\ddot{\rm a}$hler potential is 
as we have chosen it. In more complicated
situations the above formulas have to be modified accordingly. 
Also, ${\cal G}^{i \bar j}=diag((S+\bar S)^2,-1)$ and 
${\cal G}^i={\cal G}^{i \bar j}{\cal G}_{\bar j}$,  
${\cal G}^{\bar i}={\cal G}_j{{\cal G}^{j \bar i}}$. 
A simple computation then yields: 
\be Q=N-1-3(S+{\bar S})^2|{F_S\over {\cal W}}|^2. \ee 
Remembering that $S+{\bar S}={2\over g^2}\sim {\cal O}(1)$ 
and that in realistic models 
$N\sim {\cal O}(150)$
(in the second reference in \cite{CI} $N=143$),
we conclude that 
if ${|{F_S\over {\cal W}}|}\sim  {\cal O}(1)$
(this does not exclude the case 
$|{F_X\over {\cal W}}|\sim {\cal O}(1)$), using the 
condition for the vanishing of the tree level
cosmological constant 
${|{F_S\over {\cal W}}|}\simeq 3$, we get the following
condition for $Q=0$:
$N-1\simeq {36\over g^4}$ 
\footnote{This relation is rather stable.
By including in $K$ the $-3\log{(T+{\bar T})}$ term,
it changes to $N\simeq {36\over g^4}$.}.
It is amusing to notice that $N=150$ implies 
$\alpha \simeq 1/26$. In other words, in dilaton dominated models
with $V=0$ at the minimum
(the same goes through for other moduli dominated models), the one loop
cosmological constant vanishes for values of the coupling constant
remarkably close to its unification value. The dilaton $F$-term,
therefore, has to be of  
at least equal order of magnitude as the $F_X$ term. 

\section{Conclusions}

We analyzed two simple models of gauge mediated supersymmetry breaking
in the context of supergravity and in particular
we looked for supersymmetry breaking vacua with zero cosmological constant.
To be able to cancel the
tree level cosmolological constant, 
we allowed for a constant $k$ in the superpotential.
With a K$\ddot{\rm a}$hler potential of the form ($\ref{eq:Kahler}$) and
$k=0$ it is not only impossible to do the latter but also it does not
seem possible to get a weak coupling minimum $\cite{BA}$.    
\par For the GMSB model, we showed that it is not possible to have 
a supersymmetry breaking vacuum state with 
zero tree level cosmological constant, with $X$ in the desired range for
low energy gauge mediation. We can speculate what
would happen if we relaxed some of our simplifying assumptions. 
We could generalize the K$\ddot{\rm a}$hler potential by including $T$ or $U$
moduli but our conclusions 
regarding the vanishing of the cosmological constant
are unlikely to be changed, except that 
some other modulus $F$-term or a linear combination of them
would take the place of $F_S$. 
Corrections to the tree level K$\ddot{\rm a}$hler potential
would be small in the weak coupling limit and thus unlikely to change
these conclusions. Our conclusion then 
is not that these models are ruled out but that 
they require a dilaton stabilization at the string
scale so that it is frozen out of the low energy field theory dynamics.
\par 

\par For the U1MSB model, on the other hand, 
we had to assume that the $D$-terms were negligible 
in order to avoid conflict with fcnc data. 
Provided that this was the case,
we argued that there might exist (dilaton dominated) supersymmetry breaking
minima with vanishing tree level and one loop cosmological constant
with a field theoretic mechanism for dilaton stabilization.      
  
\section{Acknowledgments}
This work was supported in part by the 
United States Department of Energy under grant DE-FG02-91-ER-40672.


\begin{thebibliography}{Ref}
\bibitem{Nilles} 
H. P. Nilles Phyics Reports C110 (1984) 1,
J. P. Deredinger, L. Ib\'a\~nez, H. P. Nilles Phys. Lett. 155B (1985) 65,
M. Dine, R. Rohm, N. Seiberg, E. Witten Phys. Lett. 165B (1985) 55,
H. P. Nilles Phys. Lett. 115B (1982) 193 and 
Nucl. Phys. B217 (1983) 366. 
\bibitem{BA}
R. Brustein, hep-th/9405066,
J. A. Casas, Phys. Lett. B384 (1996) 103, 
R. Brustein, S. P. de Alwis, hep-th/0002087.
\bibitem{GuidRat} 
G. F. Guidice, R. Rattazzi hep-ph/9801271 and references therein.
\bibitem{DN}
M. Dine, A. E. Nelson Phys. Rev. D48 (1993) 1277.
\bibitem{Dimopoulos}
S. Dimopoulos, G. Dvali, R. Rattazzi, G. F. Guidice 
Nucl. Phys. B510 (1998) 12. 
\bibitem{Yanagida}T.~Yanagida,
Phys.\ Lett.\  {\bf B400}, 109 (1997)
[hep-ph/9701394].
\bibitem{BD} 
P. Bin\'etruy, E. Dudas, Phys. Lett. B389 (1996) 503,
G. Dvali, A. Pomarol Phys. Rev. Lett. 77 (1996) 3728,
N. Arkani-Hamed, M. Dine, S. Martin Phys. Lett. B431 (1998) 329.  
\bibitem{ADS}
I. Affleck, M. Dine, N. Seiberg Nucl. Phys. B241 (1984) 493,
and Nucl. Phys. B256 (1985) 557,
N. Seiberg Phys. Lett. B318 (1993) 469.
\bibitem{sugra} 
E. Cremmer, S. Ferrara, L. Girardello, A. Van Proeyen
Nucl. Phys. B212 (1983) 413.
\bibitem{ILR}
L. Ib\'a\~nez, G. G. Ross Phys. Lett. B332 (1994) 100,
N. Irges, S. Lavignac, P. Ramond Phys. Rev. D58 (1998) 035003
and references therein.
\bibitem{CI}
T. Barreiro, B. de Carlos, J.A. Casas, J.M. Moreno Phys. Lett. B445 (1998) 82, 
N. Irges 
Phys. Rev. D58 (1998) 115011,
Phys. Rev. D59 (1999) 115008. 
\bibitem{CW} 
S. Coleman, E. Weinberg Phys. Rev. D7 (1973) 1888,
S. Weinberg Phys. Rev. D7 (1973) 2887,
J. Iliopoulos, C. Itzykson, A. Martin Rev. Mod. Phys. 47 (1975) 165.
\bibitem{FerKounZwir}
S. Ferrara, C. Kounnas, F. Zwirner
Nucl. Phys. B429 (1994) 589, Erratum-ibid. B433 (1995) 255.

\end{thebibliography}
\end{document}